\documentstyle[aps,preprint,epsfig]{revtex}
\tighten

\begin{document}

\csname @twocolumnfalse\endcsname

\title{Correlations in quantum systems and branch points in the complex
plane}

\author{I Rotter\footnote{Electronic address: rotter@mpipks-dresden.mpg.de} }
\address{
 Max-Planck-Institut f\"ur Physik
komplexer Systeme, D-01187 Dresden, Germany}

\date{\today}

\maketitle

\begin{abstract}
Branch points in the complex plane are responsible for avoided level crossings
in closed and open quantum systems. They create not only an exchange of the
wave functions but also a mixing of the states of a quantum system
at high level density. 
The influence of branch points in the complex plane on the low-lying states 
of the system is small. 
\vspace*{.3cm}

\end{abstract}

\pacs{05.45.-a, 21.60.Cs, 05.30.-d 24.60.Lz}



\section{Introduction}

The statistical properties of the shell-model states of nuclei 
around $^{24}$Mg  are
studied a few years ago \cite{zelev} by using two-body forces which are 
obtained  by fitting the low-lying states of  different nuclei
of the $2s-1d$ shell. In the center of the spectra,
the generic properties are well expressed  in spite 
of the two-body character of the forces used in the calculations.
It is therefore not surprisingly that 
embedded random matrix ensembles are  relevant 
for a description of  eigenfunctions and  transition strengths
in many-body quantum systems \cite{flam,izr,kota}.
Some regularity in the  ground state region of the spectra
is obtained in recent calculations with the two-body random ensemble  
\cite{bertsch,frank,kaplan} which
could be explained in the cases considered \cite{zel,drozdz}. 

In the calculations  with the two-body random ensemble, the two-body 
character of the forces in the Hamiltonian and the Pauli principle are 
taken into account:
states with  given total angular momentum $J$ 
consist of $N$ identical Fermi-particles 
distributed among a set of single-particle states, and
interact via two-body matrix elements. 
By standard shell-model (SM) techniques, the  Hamiltonian matrix
is calculated in terms of the independent two-body matrix elements
which are taken  at random 
from a given distribution. The number of independent two-body matrix elements 
is small as compared to the total number of matrix elements.
Due to its nearness to the SM, the two-body random ensemble
is expected to be more appropriate than the Gaussian orthogonal ensemble
for a study of nuclear spectra \cite{tbre1,tbre2}.
The level density  calculated with a two-body random ensemble  is, 
indeed, more realistic  than
that of the Gaussian orthogonal ensemble. It is nearly Gaussian  in contrast
with the  Wigner's semi-circle law, that holds for the 
Gaussian orthogonal ensemble in the limit 
of a large number of states.

In Ref. \cite{kaplan}, the spectrum of $N$  fermions  distributed
over $M$ orbitals and the structure of the eigenstates is calculated with a
two-body random ensemble  by using the methods of scar theory.
At the edges of the spectrum, significant deviations 
from random matrix theory  expectations are found.
The deviations are much smaller when
a spin is added to the Hamiltonian (as in nuclear SM 
calculations). Due to the spin,
the single-particle states split and
the many-particle  SM states may cross.
The crossing of any interacting discrete states is, however, not free but 
avoided. 
Thus, the results of the calculations \cite{kaplan} with and without spin  
differ by the number of avoided crossings of their levels. 

It is the aim of the present paper to study the relation between avoided 
crossings of levels  and the mixing of their wave functions  in detail.
The avoided  crossings of the {\it many-particle} levels are 
a property of the whole system consisting of
{\it many} particles. The   correlations induced by them 
are a second-order effect (via the continuum)
as compared to those induced directly by the
two-body forces. Nevertheless, they may play an important role,
especially at high level density.

In section 2, the relation between branch points and avoided level crossings
is discussed by means of a simple example.
The correlations induced by the branch points in the complex
plane are considered in section 3. 
Numerical examples are shown and discussed in section 4 and
the results obtained are summarized in the last section.

\section{Branch points in the complex plane and avoided level crossings}

To every avoided  crossing of discrete states, the corresponding crossing
point in the complex plane can be found. 
This can be seen by considering the matrix
\begin{eqnarray}
{\cal{H}}= \left(
\begin{array}{cc}
e_1(a) & v\\
v & e_2(a)
\end{array} 
\right) 
- \frac{i}{2}
\left(
\begin{array}{cc}
c_1(a) & 0\\
0 & c_2(a)
\end{array} 
\right) 
\label{eq:matr1}
\end{eqnarray}
whose eigenvalues are
\begin{eqnarray}
{\cal E}_{i,j} \equiv
E_{i,j} - \frac{i}{2} \; C_{i,j} = \frac{\varepsilon_1 + \varepsilon_2}{2}
\pm \sqrt{(\varepsilon_1 - \varepsilon_2)^2 + 4 v^2} 
\label{eq:matr1a}
\end{eqnarray}
with $i,j = \pm$.
Further, $\varepsilon_k \equiv e_k - i/2\; c_k \; (k=1,2) $ 
is the energy
$e_k$ of the  state $k$ continued by $1/2\; c_k$ into the complex
plane. The states with energies $e_k$  are  assumed to be 
unperturbed by any interaction 
with other states. Their energies may be traced as a function of
a certain parameter $a$. Diagonalizing ${\cal{H}}$ gives the energies 
${\cal E}_{i,j}$ of the eigenstates of  ${\cal{H}}$ 
in which the interaction $v$ between the
two basic states $1$ and $2$ is taken into account. 
 
The eigenvalue equation (\ref{eq:matr1a}) shows the following.
When the  non-diagonal matrix elements $v$  (which describe the interaction 
between the two states $k=1,2$) are real and different from zero, 
the two states $i,j = \pm$
can cross, as a function of the parameter $a$, only in the complex plane.
The crossing point $a=a^{\rm cr}$ is defined by 
${\cal E}_{+}(a^{\rm cr}) = {\cal E}_{-}(a^{\rm cr})$
and 
\begin{eqnarray}
F^{\rm cr} = \{\varepsilon_1(a^{\rm cr}) - \varepsilon_2(a^{\rm cr})\}^2 + 
4 v^2 = 0 \; .
\label{eq:matr1b}
\end{eqnarray}
This crossing point  is a branch point in the complex plane
according to Eq. (\ref{eq:matr1a}). It lies at $\{\varepsilon_1(a^{\rm cr}) +
\varepsilon_2(a^{\rm cr})\}/2$.  

 The non-trivial 
mathematical properties of these branch points 
in the complex plane as well as their
relation to the higher-order poles of the $S$ matrix 
are known for a long time, e.g. \cite{newton}. 
The eigenfunctions $\tilde \Phi_R$ of ${\cal H}$ are bi-orthogonal.
They are normalized  \cite{pegoro} according to 
\begin{eqnarray}
(\tilde \Phi_R )^2 & \equiv &
\langle \tilde\Phi_{R'}^* | \tilde\Phi_{R} \rangle  =  \delta_{RR'} 
\label{eq:bio1} 
\end{eqnarray}
and therefore
\begin{eqnarray}
|\tilde \Phi_R |^2 & \equiv & \langle \tilde\Phi_R | \tilde\Phi_{R} \rangle  
\geq  1  \; ;  \quad
| \langle \tilde\Phi_{R' \ne R} | \tilde\Phi_{R} \rangle| 
\equiv |{\rm Im} \langle \tilde\Phi_{R' \ne R} | \tilde\Phi_{R} \rangle| 
 \geq  0 \; .
\label{eq:bio2} 
\end{eqnarray}
Approaching a double pole,
\begin{eqnarray}
\tilde\Phi_1^{\rm bp} 
\to \pm \; i \; \tilde\Phi_2^{\rm bp}
\label{eq:bio3} 
\end{eqnarray}
see \cite{marost}, and
\begin{eqnarray}
|\tilde \Phi_R^{\rm bp} |^2 
\to \infty   \; ;  \quad    | \langle \tilde\Phi_{R' \ne R}^{\rm bp}| 
\tilde\Phi_{R}^{\rm bp} \rangle| \to \infty \; ,
\label{eq:bio4} 
\end{eqnarray}
see \cite{mudiisro}.
At an avoided crossing in the complex plane and in its  neighborhood, 
the difference between
$|\tilde \Phi_R |^2 $ and $(\tilde \Phi_R )^2$
is large but finite. Numerical examples are given in 
\cite{pegoro,mudiisro}. 
 The $S$ matrix  contains 
$\langle\tilde\Phi_R^*|\tilde\Phi_{R}\rangle$ but not 
$\langle \tilde\Phi_R | \tilde\Phi_{R} \rangle$,
i.e. it behaves smoothly also when the  bi-orthogonality
is large. This holds also at a double pole of the $S$ matrix where  two
resonance states cross \cite{mudiisro}.

The influence of  avoided level crossings 
on the dynamics of a system under adiabatic conditions 
and on the correlations between the different states   
has been studied in atoms \cite{marost,kylstra1,story} 
and open quantum billiards \cite{ropepise,perostba}.
In these systems, it is possible to trace 
the turn-over of the avoided crossings into a real crossing
(corresponding to a double pole of the $S$ matrix) by tuning the system
parameters.
All the results obtained show that  branch points in the complex 
plane may have an important influence on
the spectroscopic properties of a quantum system
although their number  is of measure zero.

\section{Correlations induced by branch points in the complex plane}

The  correlations between  states are expressed usually 
by the mixing of their wave functions
\begin{eqnarray}
\tilde\Phi_R = \sum_i b_{R,i} \Phi_{i}^0 
\label{eq:smwf} 
\end{eqnarray}
 where the $\{\Phi_{i}^0\}$ are the eigenfunctions of 
the Hamilton operator ${\cal H}_0$ of the unperturbed system.
In the SM calculations, the functions $\{\Phi_{i}^0\}$ are the Slater
determinants. They do not contain any residual forces and 
represent therefore a natural basis set for the SM wave functions 
$\Phi_R^{\rm SM}$. In the case of an open quantum system, the functions 
 $\{\Phi_{i}^0\}$ are the wave functions of the corresponding closed system. 
In the  continuum shell model, these are the SM wave functions
$\{\Phi_R^{\rm SM}\}$.
In the  case of the matrix  (\ref{eq:matr1a}),  ${\cal H}_0$
is defined by    $v=0$ and the $\{\Phi_{i}^0\}$ are the eigenfunctions of 
${\cal H}_0$. 

The  Schr\"odinger equation of an open quantum system with the Hamiltonian
${\cal H}$ can be written as  
\begin{eqnarray}
({\cal H}_0  -  \tilde {\cal E}_R) |\tilde\Phi_R\rangle & = & - \; W \;
 |\tilde\Phi_R\rangle 
 \nonumber \\
& = & - \sum_{R'} \langle \tilde\Phi_{R'}| W |\tilde\Phi_R\rangle
 \sum_{R''} \langle  \tilde\Phi_{R'}| \tilde\Phi_{R''}\rangle 
|\tilde\Phi_{R''}\rangle 
\label{eq:nls} 
\end{eqnarray}
where the operator $W$  describes the interaction of the two states 
$R$ and $R'$ via the environment.
In the case of the continuum shell model, this is
the interaction via the continuum which is complex, as a rule 
\cite{ro91,drokplro}.
The source term  (rhs. of Eq. (\ref{eq:nls})) 
contains the $ \langle \tilde \Phi_R |\tilde\Phi_{R'}\rangle $. 
Thus, the  mixing  of the wave functions via the continuum is related 
to the bi-orthogonality of the eigenfunctions of ${\cal H}$,
Eq. (\ref{eq:bio2}).  Numerical examples 
for the  bi-orthogonality, given in e.g. Refs.
\cite{pegoro,mudiisro,ropepise,slaving}, show that the 
standard relations  $ \langle \tilde \Phi_R |\tilde\Phi_{R'}\rangle
= \delta_{RR'} $ are not fulfilled neither at the branch point in the complex
plane   nor  at an avoided level crossing, see Eq. (\ref{eq:bio2}).
Thus the source term of Eq. (\ref{eq:nls}) contains  
terms being non-linear in the wave
functions  $\tilde \Phi_R$ when  the quantum system is open and 
the $\tilde \Phi_{R}$ are complex.

SM calculations are performed with an Hermitian Hamilton operator 
${\cal H}^{\rm SM}$. The eigenvalues and eigenfunctions are real. 
It seems therefore, 
that the source term in Eq. (\ref{eq:nls})
 can not play any role in the SM calculations. 
This is, however, not true as will be shown in the 
following. The reason for this unexpected behaviour is, above all,
the analyticity of the wave functions. 
The eigenvalues lying at negative energy show, as a function of tuning the 
system parameters, the same motion in the complex plane as those at positive 
energy.  

In the following, the motion of the  eigenvalues by varying the system 
parameters will be considered in a language more fitting into the 
concept of the SM. As a matter of fact, SM calculations provide 
not only the eigenvalues $E_i$ of the SM states but also the 
eigenfunctions $\Phi_i$. Knowing the SM wave functions $\Phi_i^A$ of
the system consisting of $A$ particles and 
those of the system with one particle less, the amplitudes $\langle \Phi_i^A |
 \Phi_c^{A-1}\rangle $ of the spectroscopic factors can be calculated. 
Here, $\Phi_c^{A-1}$ is the wave function of a well-defined state of the 
residual system having one bound particle less than the initial system. These 
amplitudes depend on the quantum numbers of the separated particle. By 
multiplying the spectroscopic factor with an energy dependent common 
penetration factor, one gets the probability for the transition of the 
initial state $i$ (with $A$ bound particles) to the decay channel $c$ 
(with one particle in relative motion to a certain state of
the  system with $A-1$ bound particles). The 
separated particle may either be transferred (at negative energy) or emitted 
(at positive energy). In the last case, the product of spectroscopic factor 
and penetration factor is identified with the partial decay width 
$(\gamma_i^c)^2$ of the state $i$ into the channel $c$. Further, $\Gamma_i = 
\sum_c (\gamma_i^c)^2$ is the width of an isolated state. Thus, the most 
characteristic 
part of the width $\Gamma_i$ of an isolated state is involved in the 
spectroscopic 
factor  calculated from  SM wave functions.
 
As a result, the SM eigenvalues and eigenfunctions 
of the two bound  systems  with $A$ and $A-1$ particles,
respectively, contain 
all information on the motion of the 
complex values 
$E_i - \frac{i}{2}G_i$ (where $G_i$ is proportional to the spectroscopic
factor)  
by varying the system parameters, although both
Hamiltonians ${\cal H}_0^A$ and ${\cal H}_0^{A-1}$ are Hermitian and 
their eigenvalues are real. The accompanying
motion of the energies $E_i$ is considered in many studies and avoided level 
crossings have been observed. The  accompanying  motion of the 
spectroscopic factors, however, is  not considered up to now.

\section{Numerical results}

\begin{figure}
\begin{minipage}[tl]{7.5cm}
\hspace*{-.8cm}
\psfig{file=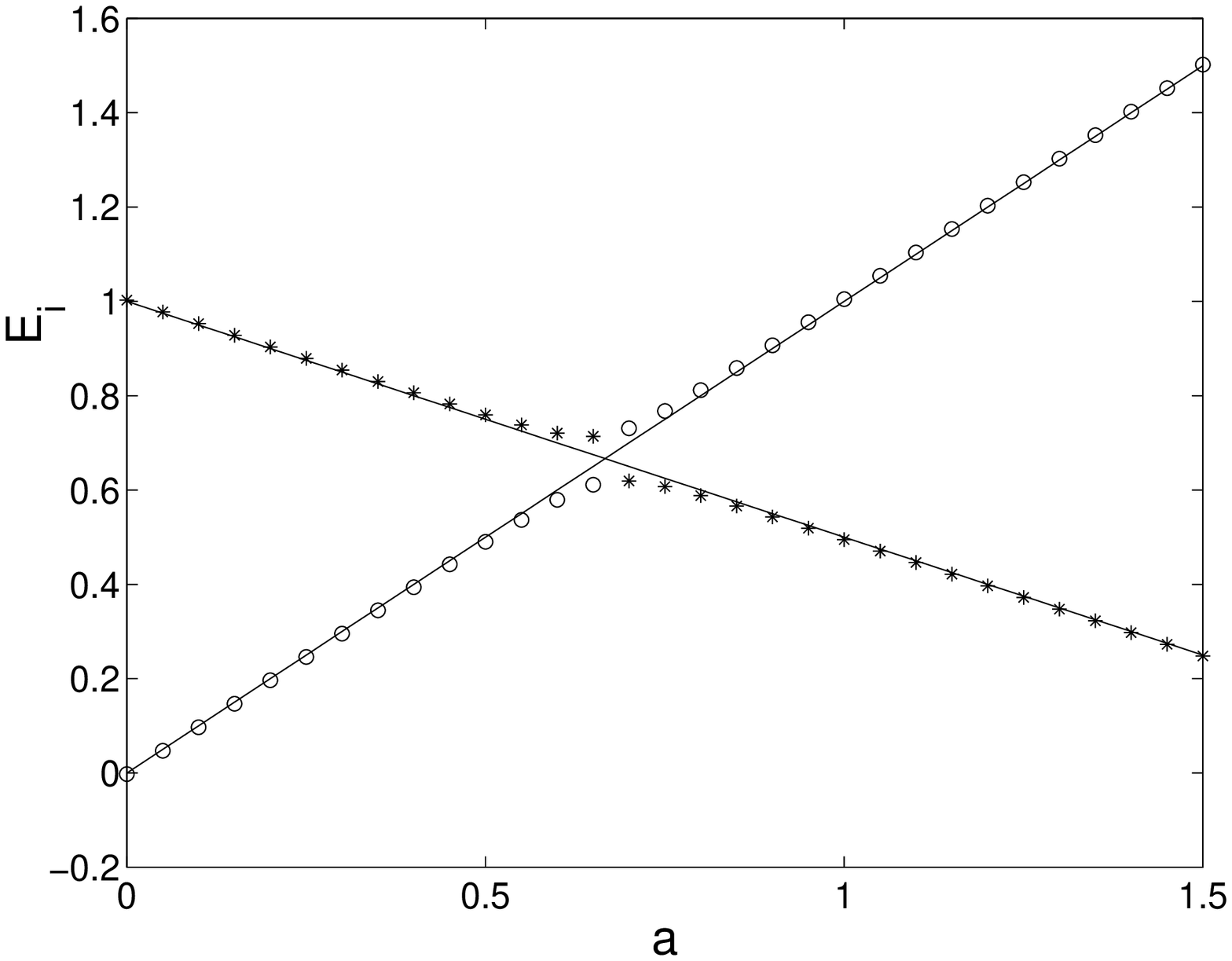,width=7.5cm}
\end{minipage}
\begin{minipage}[tr]{7.5cm}
\hspace*{-1.5cm}
\psfig{file=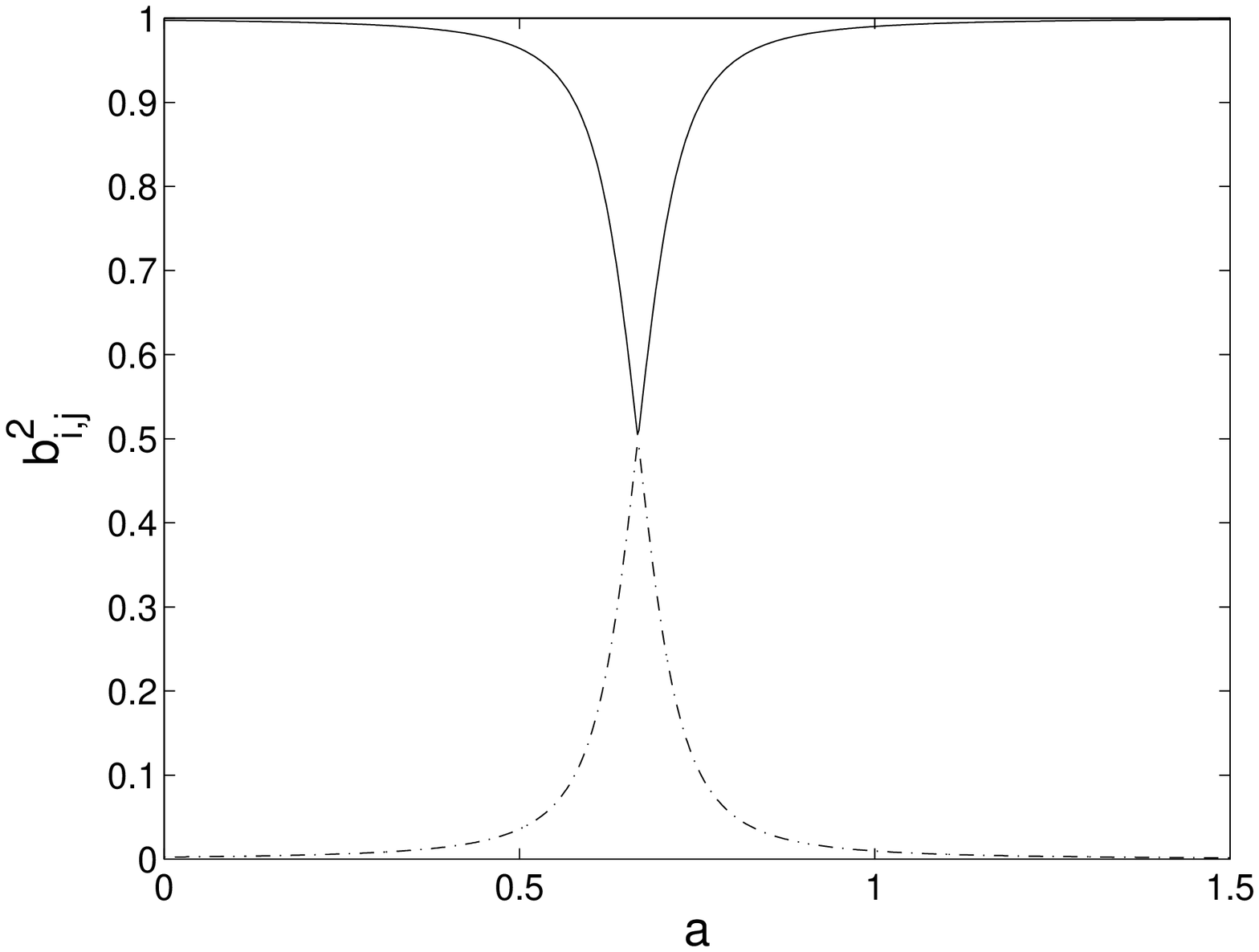,width=7.5cm}
\end{minipage}
\begin{minipage}[ml]{7.5cm}
\psfig{file=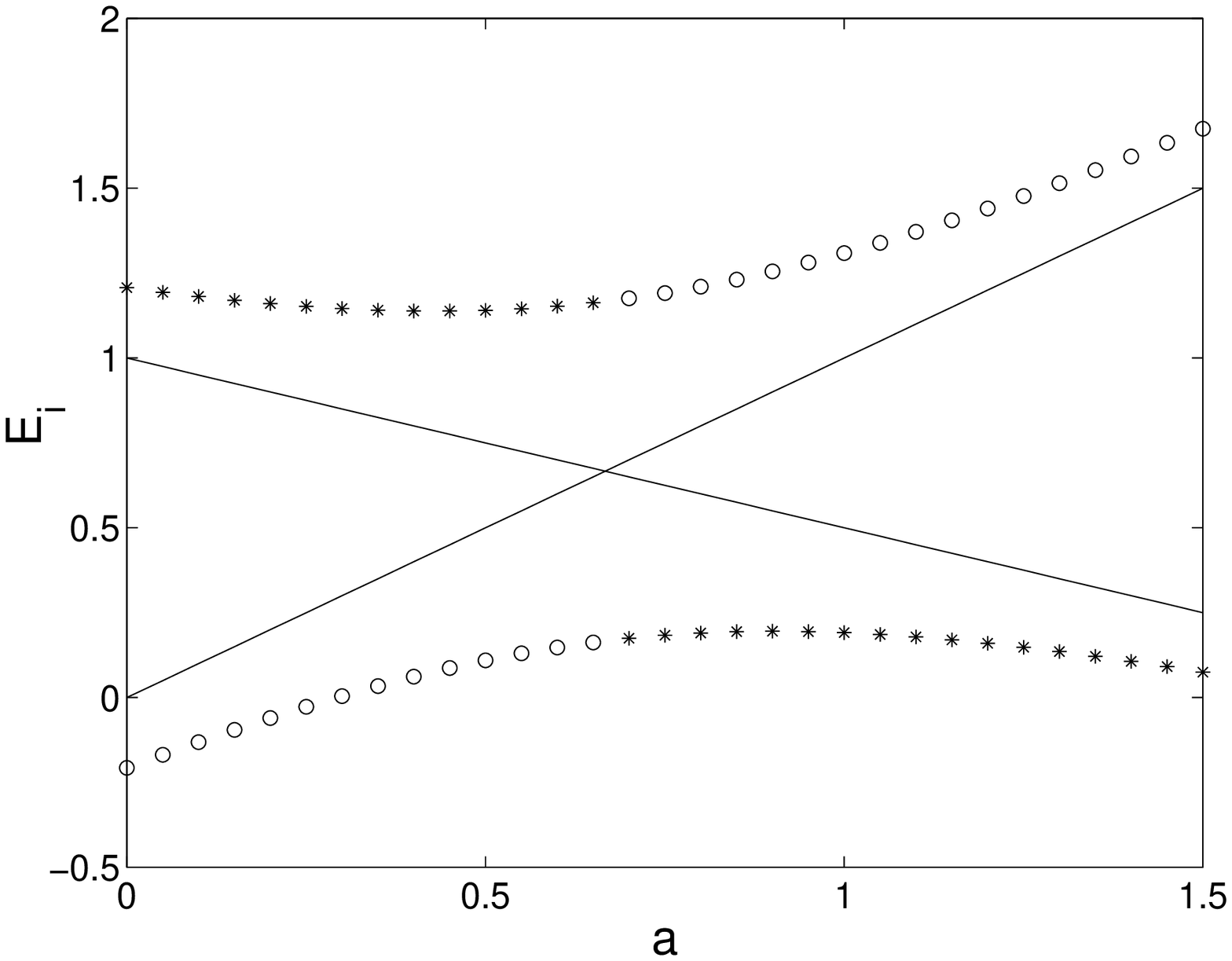,width=7.5cm}
\end{minipage}
\begin{minipage}[mr]{7.5cm}
\hspace*{-1.5cm}
\psfig{file=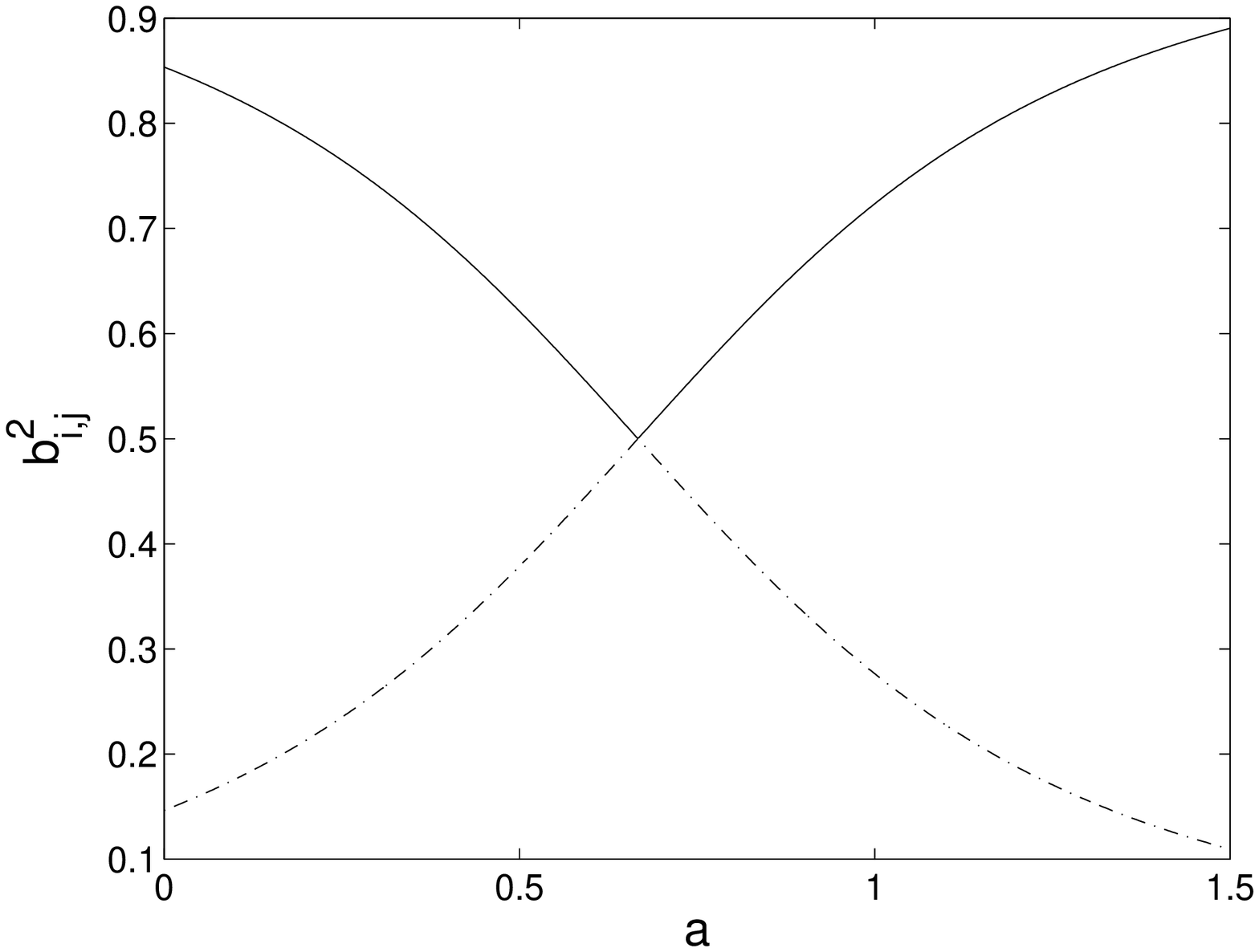,width=7.5cm}
\end{minipage}
\begin{minipage}[bl]{7.5cm}
\psfig{file=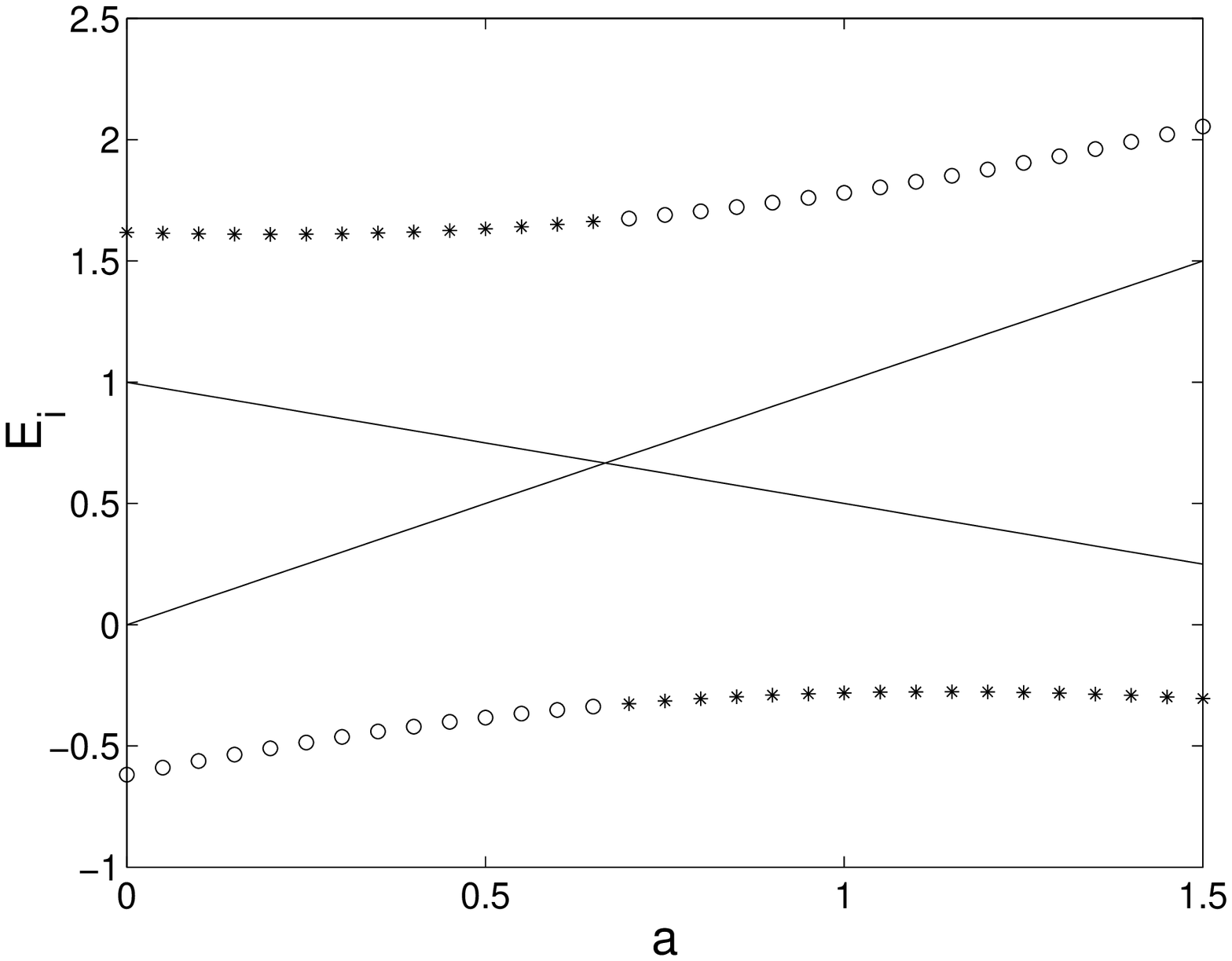,width=7.5cm}
\end{minipage}
\begin{minipage}[br]{7.5cm}
\psfig{file=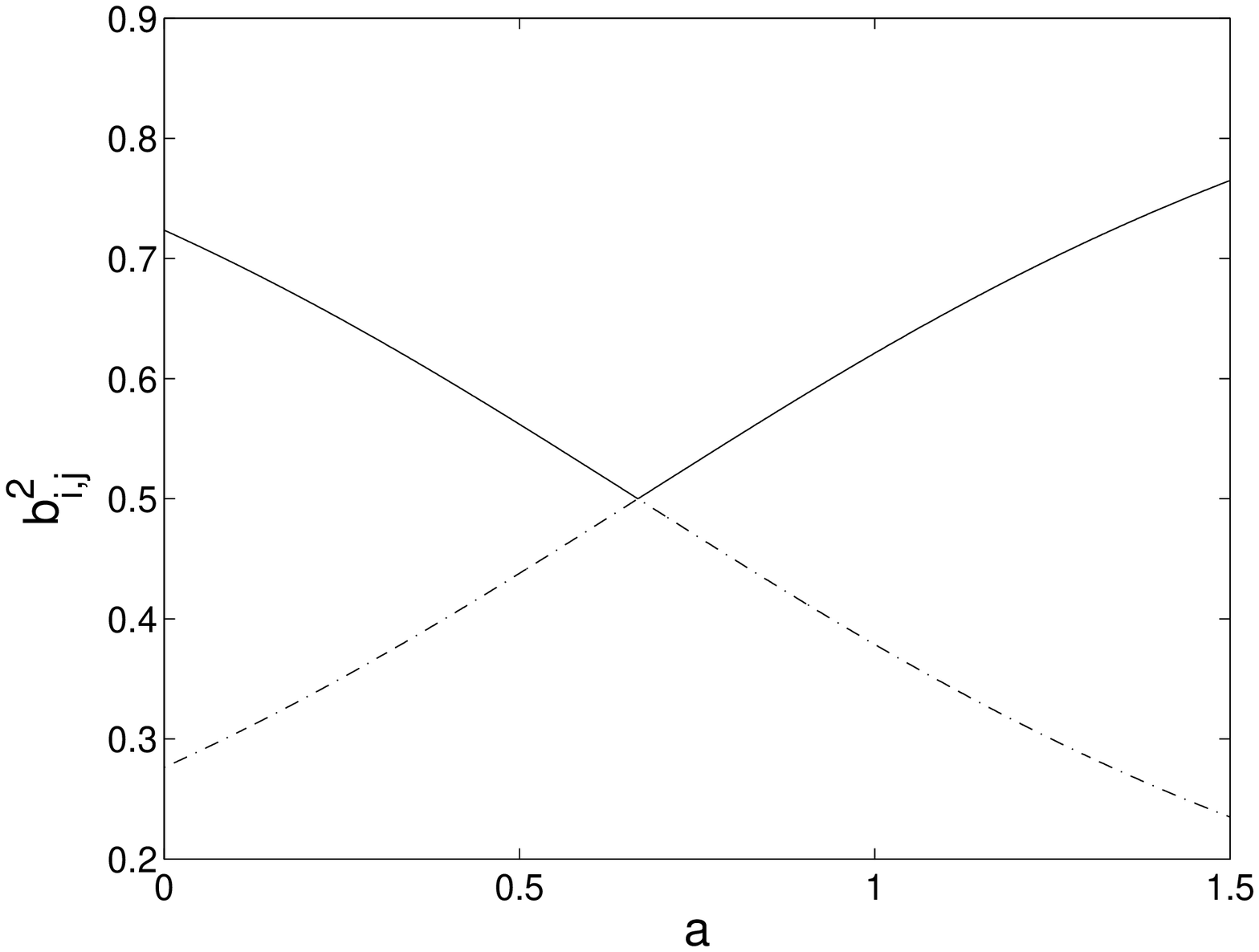,width=7.5cm}
\end{minipage}
\vspace*{1cm}
\caption{The energies $E_{i}$ (left)
 and mixing coefficients $b_{i,j}^2$ (right) for  two states 
$i=1, 2$      as a function of $a$ for
$v=0.05$ (top), 0.5 (middle) and 1.0 (bottom).
The circles ($i=1$) and stars ($i=2$) on the left-hand side show the 
energies of the two states at $v \ne 0$ and the 
full lines show their energies  for $v=0$ (which cross at $a = a^{\rm cr}
=2/3$).
The full lines on the right-hand side are the coefficients $b_{i,j=i}^2$ while 
the dashed lines are the $b_{i,j \ne i}^2$. Note the different scales 
(top, middle and bottom)
of the  $E_i$ and   $b_{i,j}^2$ axes.  }
\label{fig:energ}
\end{figure}

In order to illustrate the motion of the energies and spectroscopic factors
as a function of a certain parameter, let us consider the 
matrix ${\cal H}$, Eq. (\ref{eq:matr1}), 
with $ e_1 =  1-  a/2   ; \; e_2 = a    ;
\;\; c_1 = c_2 = 0$ and $v$ real.
By means of the parameter $a$,  crossing or avoided crossing of the two 
states can be traced under different conditions, e.g. for different values of 
$v$. The results show
crossing for $v=0$ and avoided crossing 
for all $v\ne 0$  at the critical value $a^{\rm cr} = 2/3$. 
The influence of $v$ on the eigenvalues is small as long as 
$v$ is small: the typical avoided-crossing behaviour can be seen
(Landau-Zener effect, figure \ref{fig:energ} top left). 
The two states are exchanged, what is 
surely without any interest for statistical considerations. 
For larger $v$, the avoided crossing of the eigenvalues as a function of 
$a$ is difficult to see in the trajectories: the eigenvalues are weakly 
dependent on $a$. The comparison with the eigenvalues for 
$v=0$ shows, however, the  changes of the 
trajectories in the neighborhood 
of   $a^{\rm cr}$. The wave functions are exchanged
also in this case at  $a = a^{\rm cr}$.
The wave functions are mixed strongly near  $a^{\rm cr}$ for all $v$ 
as can be seen from  the right-hand side of
figure \ref{fig:energ}. Here the
coefficients $b_{i,j}^2$ of the representation $\Phi_i = \sum b_{i,j}
\Phi^0_j$, Eq. (\ref{eq:smwf}), 
 for $v=0.05, \; 0.5$ and 1 are shown ($\Phi^0_j$ are the wave functions 
at $v=0$). 

While the wave functions are mixed only in a very small region around
the critical value $a^{\rm cr}$ for small $v$, they are mixed also 
at large distances $|a- a^{\rm cr}|$  
when $v$ is larger (figure  \ref{fig:energ}). 
At high level density, avoided level
crossings with other states  appear, as a rule,  before   $b_{i,i} \to 1 \; ,
\; b_{i,j\ne i} \to 0 $ is reached. As a consequence, the $b_{i,j}$ 
contain components from more than just the two crossing levels. 

\begin{figure}
\begin{minipage}[tl]{7.5cm}
\hspace*{-.6cm}
\psfig{file=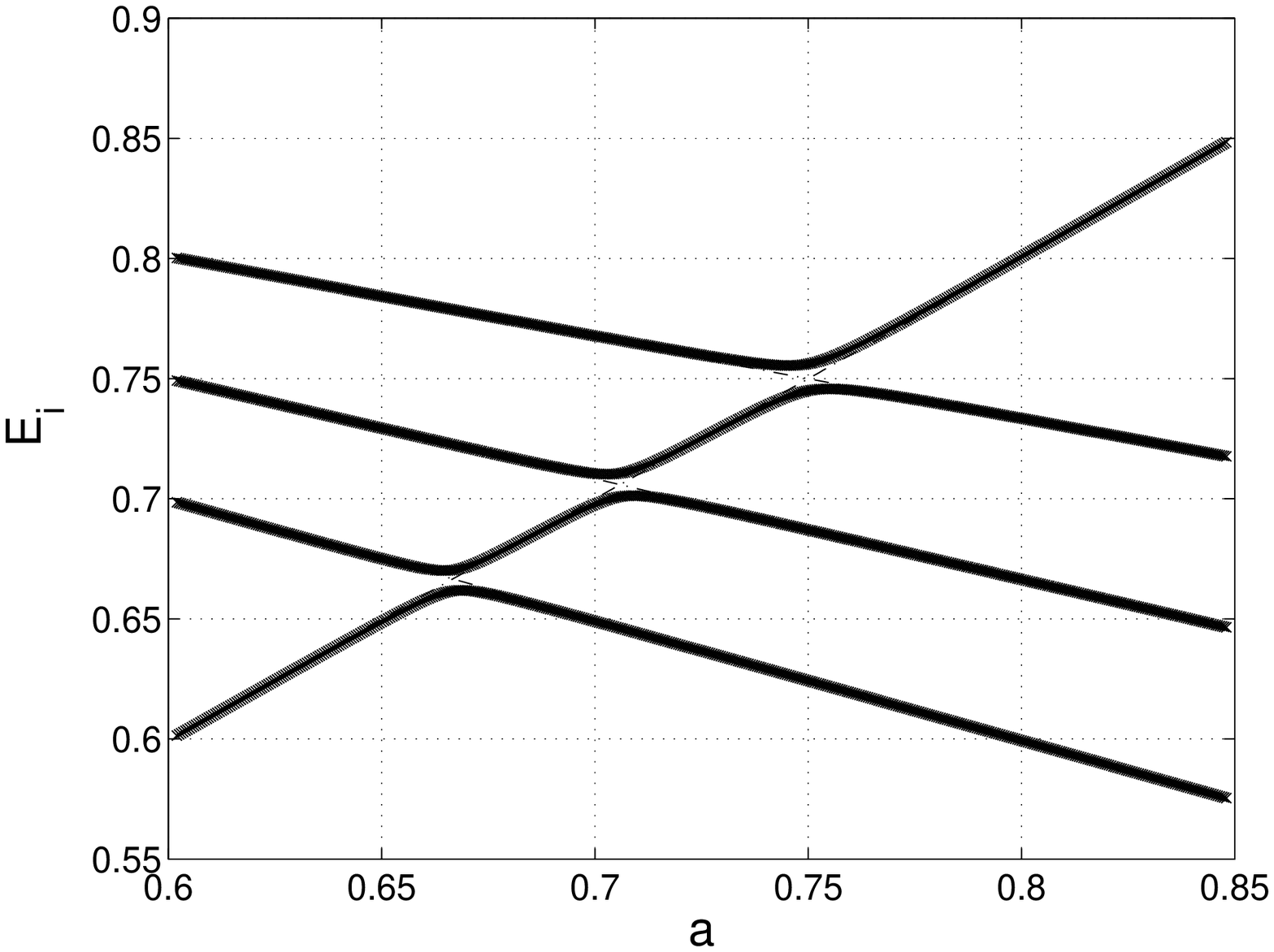,width=7.5cm}
\end{minipage}
\begin{minipage}[tr]{7.5cm}
\hspace*{-1.5cm}
\psfig{file=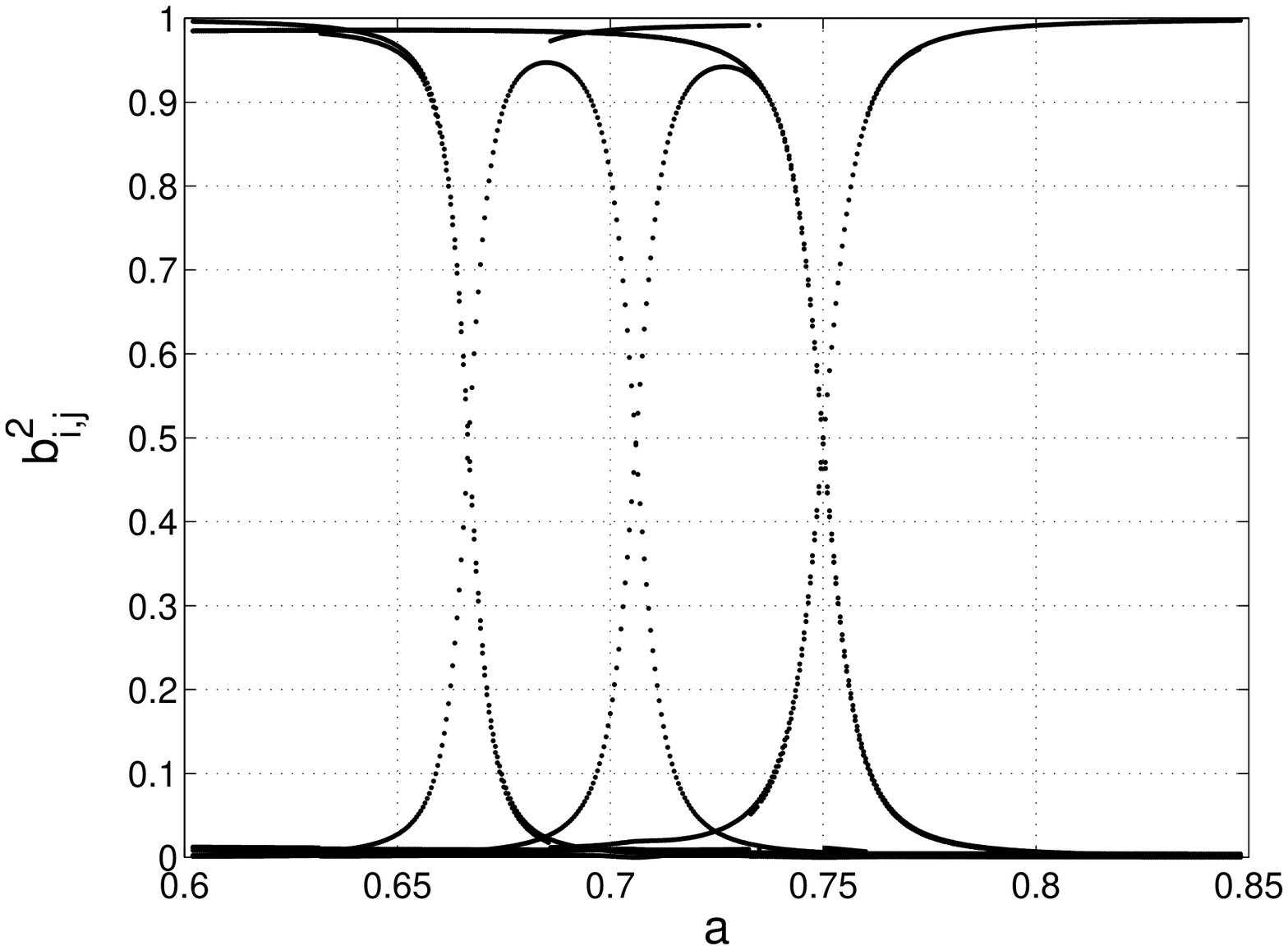,width=7.5cm}
\end{minipage}
\begin{minipage}[bl]{7.5cm}
\hspace*{.2cm}
\psfig{file=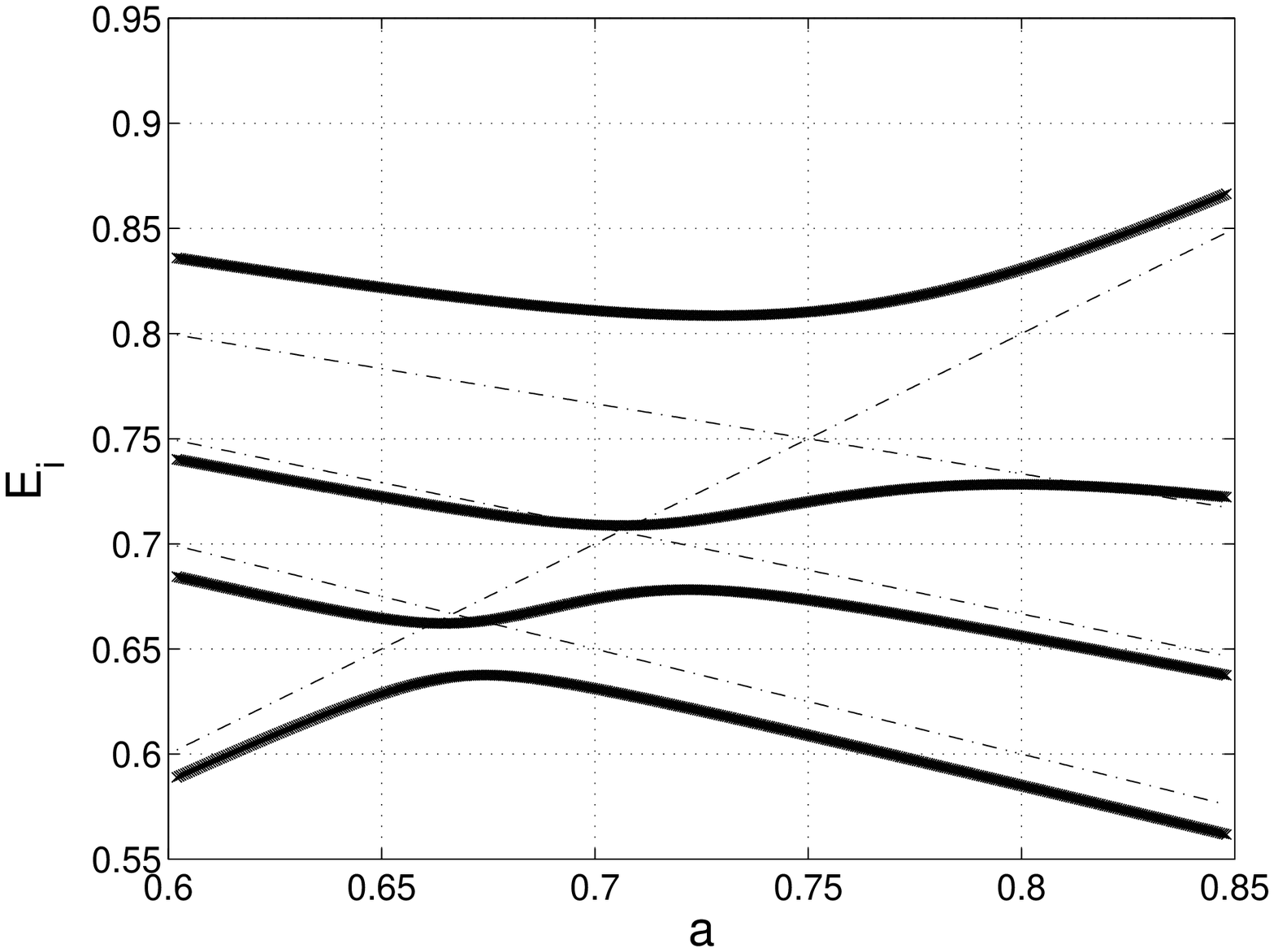,width=7.5cm}
\end{minipage}
\begin{minipage}[br]{7.5cm}
\psfig{file=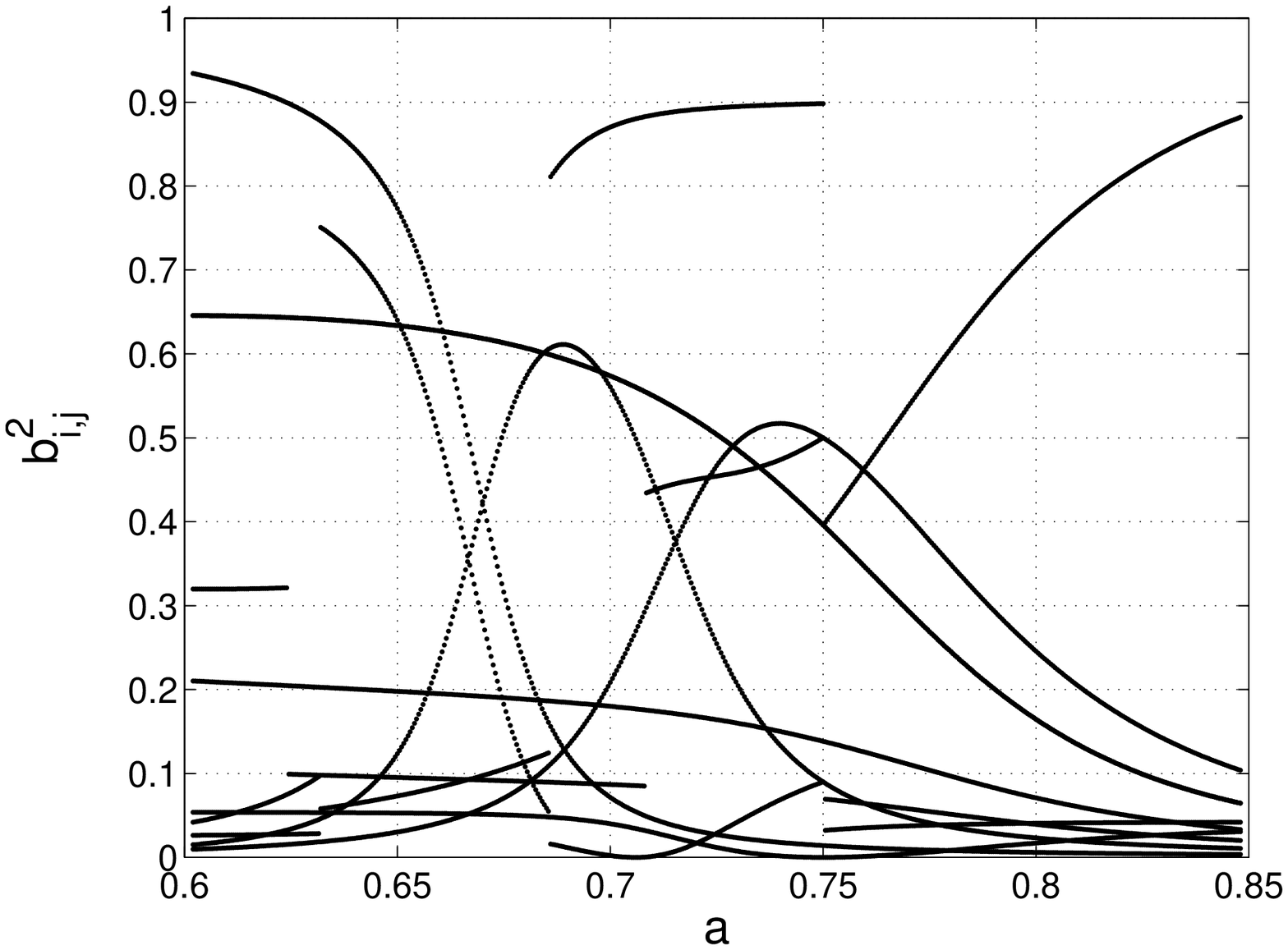,width=7.5cm}
\end{minipage}
\vspace*{1cm}
\caption{The energies $E_{i}$ (left)
 and mixing coefficients $b_{i,j}^2$ (right) for four states 
     as a function of $a$ at
$v=0.005$ (top) and 0.03 (bottom) for all coupling matrix elements.
The full lines on the left-hand side show the 
energies of the  states at $v \ne 0$ and the 
dash-dotted lines show their energies  at $v=0$.
It is $b_{i,j=i}^2 \ge b_{i,j\ne i}^2 $
($\; b_{i,j=i}^2 = b_{i,j\ne i}^2 =0.5$ at $a_i^{\rm cr}$)
 for all $i$ at small $v$ 
(top right) as in figure \ref{fig:energ}. At large $v$, every state is mixed
 with all the other ones and an unequivocal identification of the 
$i$ and $j$ in the mixing
coefficients  $b_{i,j}^2$ is impossible. 
}
\label{fig:energ2}
\end{figure}

For illustration, the energies and wave functions of
four states are calculated from
 \begin{eqnarray}
{\cal H}^{(4)} =
 \left(
\begin{array}{cccc}
 e_1(a) &  v & v & v \\
  v & e_2(a) & v & v \\
  v & v & e_3(a) & v \\
 v & v &  v & e_4(a)\\
\end{array}
\right)  
\label{eq:matr4}
\end{eqnarray}
 and shown in figure \ref{fig:energ2} 
for the two values $v=0.005$ and 0.03
($e_1=1-a/3; \; e_2= 1-(5/12) \; a; \; e_3=1-a/2; \; e_4=a $). 
As long as $v$ is small, the influence of the branch points 
 consists mainly in an exchange of the corresponding two states and
of the mixing of just these two wave functions  in a region  
$|a- a^{\rm cr}|$  which is well separated from other analogous regions
around other critical values.
At larger $v$, however, the states are exchanged 
and  mixed with {\it all} the other ones since the different regions
 $|a- a^{\rm cr}_k|$, corresponding to different branching points, overlap.

In the special case considered, the regions $|a- a^{\rm cr}_k|$ in which
$b_{i,i} \to 1 \; ,
\; b_{i,j\ne i} \to 0 $ is reached, are well separated for the different
critical values $a^{\rm cr}_k$ as long as
$v<0.02$. The overlap of the different regions  $|a- a^{\rm cr}_k|$
 starts at  $v\approx 0.02$. Here, 
the mixing of the wave function of every state with those of 
all the other ones starts. It changes  smoothly with increasing $v > 0.03$. 
Characteristic of this regime is that an unequivocal identification
of the  $i$ and $j$ in the mixing coefficients  $b_{i,j}^2$
is impossible. In this manner, correlations between  the 
wave functions of all the states may be created by the avoided crossings,
more exactly, by the branch points in the complex plane (expressed
mathematically by the source term in the Schr\"odinger equation 
(\ref{eq:nls})).

Due to the mixing, the 
wave functions of the  states become more similar and the differences of 
their spectroscopic factors in relation to a certain channel
become less. That means, the level repulsion 
along the real axis is accompanied by an adjustment  of the 
spectroscopic factors. This 
result coincides fully with the results obtained from a study of the motion 
of the poles of the $S$ matrix in the neighborhood of an avoided crossing
in the complex plane. Here, level repulsion along the real axis is accompanied
by an adjustment of the widths \cite{marost,ropepise}. 

The results show that the energies of the states are almost 
independent of $a$ and $v$, 
if $v$ is sufficiently large or (and) the level density is high. 
That means, the system is stabilized against (real) 
distortions of any type. This is true also for the distortions 
introduced by the real part of the coupling to the continuum (principal value
integral) which does not vanish, generally  \cite{drokplro}. 
In any case, the states have lost 
their individual SM properties by spreading their 
spectroscopic features. This spreading is  caused by  
the source term of Eq. (\ref{eq:nls})
which continues analytically into the function space of the discrete 
states. It does no longer show features characteristic of the special type
of the two-body forces used in the calculation.

It follows immediately, that the states in the centre of a 
SM spectrum show   features caused by the branch points in the complex plane,
i.e. by the non-linear terms in the Schr\"odinger
equation. The states at the 
border of the spectrum are less influenced by avoided 
crossings with other states. This is true especially for the lowest 
states which may shift downwards  
by crossing  only a few (or no) other states.

\section{Summary}

In this paper, the influence of
branch points in the complex plane onto the 
correlations between the states of a quantum system is studied. As
a quantitative measure of
the correlations,  the degree of mixing of the wave functions, i.e. the values
$b_{R,i}$ in the representation (\ref{eq:smwf}) is used. The 
results are the following.
\begin{itemize}
\item[--]
The exchange of two discrete levels which avoid crossing can be traced back to
the existence of a branch point in the complex plane. Here, the levels have 
equal energies and widths, their wave functions are linearly dependent, and 
the Schr\"odinger
equation contains non-linear terms.

\item[--]
The correlations induced by the branch
points in the complex plane appear between the discrete states of a closed
system in the same manner as between the resonance states of an open system.

\item[--]
The number of avoided level crossings  at high level density 
is much larger than usually assumed. Most important correlations 
are introduced even by those avoided crossings 
which are difficult to identify  at first sight.

\end{itemize}

It follows further from these results that the correlations induced by the
branch points in the complex plane are the larger the higher the level density
is. At the  edges of the spectrum,  branch points have
only a small influence  on the spectrum. The properties of these states are
therefore determined, at least partly, by two-body forces.

\vspace{1cm}
{\bf Acknowledgments}
Valuable discussions with
J. Flores, M. M\"uller and T.H. Seligman  at the Centro
Internacional  de Ciencias, Cuernavaca, Mexico, and with E. Persson
are gratefully acknowledged.

\newpage

\end{document}